\documentclass[sigconf]{acmart}

\usepackage{booktabs} 

\usepackage{tikz}
\def\checkmark{\tikz\fill[scale=0.4](0,.35) -- (.25,0) -- (1,.7) -- (.25,.15) -- cycle;}

\begin{document}

\copyrightyear{2017} 
\acmYear{2017} 
\setcopyright{rightsretained}
\acmConference{PEARC17}{July 09-13, 2017}{New Orleans, TX, USA}\acmPrice{15.00}\acmDOI{10.1145/3093338.3093363}
\acmISBN{978-1-4503-5272-7/17/07}


\title{Data Access for LIGO on the OSG}

\author{Derek Weitzel, Brian Bockelman}
\affiliation{%
  \institution{University of Nebraska - Lincoln}
  \streetaddress{1101 T St.}
  \city{Lincoln} 
  \state{Nebraska} 
  \postcode{68588}
}
\email{dweitzel@unl.edu, bbockelm@cse.unl.edu}

\author{Duncan A. Brown}
\affiliation{
  \institution{Syracuse University}
  \streetaddress{Department of Physics}
  \city{Syracuse} 
  \state{New York} 
  \postcode{13244}
}
\email{duncan.brown@ligo.org}


\author{Peter Couvares}
\affiliation{
  \institution{California Institute of Technology}
  \streetaddress{1200 E. California Blvd.}
  \city{Pasadena} 
  \state{California} 
  \postcode{91125}
}
\email{peter.couvares@ligo.org}

\author{Frank W{\"u}rthwein, Edgar Fajardo Hernandez}
\affiliation{
  \institution{University of California San Diego}
  \streetaddress{9500 Gilman Drive}
  \city{La Jolla} 
  \state{California} 
  \postcode{92093}
}
\email{fkw@ucsd.edu, emfajard@ucsd.edu}

\renewcommand{\shortauthors}{D. Weitzel, et. al.}

\begin{abstract}

During 2015 and 2016, the Laser Interferometer Gravitational-Wave 
Observatory (LIGO) conducted a three-month observing campaign. These observations delivered the first direct
detection of gravitational waves from binary black hole mergers.  To search for these signals, the LIGO Scientific Collaboration uses the PyCBC search pipeline.  To deliver science results
in a timely manner, LIGO collaborated with the Open Science Grid (OSG) to
distribute the required computation across a series of dedicated, 
opportunistic, and allocated resources.  To deliver the petabytes
necessary for such a large-scale computation, our team deployed a
distributed data access infrastructure based on the XRootD server suite
and the CernVM File System (CVMFS).  This data access strategy grew from
simply accessing remote storage to a POSIX-based interface underpinned by
distributed, secure caches across the OSG.

\end{abstract}


\maketitle

\section{Introduction}

The Open Science Grid (OSG) \cite{pordes2007open} is a ``national, distributed computing partnership for data-intensive research;'' it provides a fabric of services for achieving Distributing High Throughput Computing (DHTC) \cite{livny1997mechanisms} across dozens of computational facilities.  OSG excels at allowing users to distribute their loosely-coupled batch jobs on a wide set of facilities and infrastructures.

The LIGO Scientific Collaboration's PyCBC search for gravitataional waves \cite{usman2016pycbc} is an archetypal OSG application from the computational point-of-view.  PyCBC analyses LIGO data in two-week blocks, requiring hundreds of thousands of individual HTCondor batch jobs, each lasting at least an hour.  There is enough parallelism to consume tens of thousands of cores simultaneously and no stringent latency requirements for completion.

However, the PyCBC pipeline requires terabytes of non-public input data; throughout the analysis, the data may be read up to 200 times.  Accordingly, the PyCBC pipeline was historically always run at sites with a full copy of the LIGO data on a shared filesystem.  To run PyCBC on two dozen opportunistic OSG sites would require the LIGO team to maintain two dozen full copies of their input data, an operationally daunting - if not impossible - task for small sites that may only be able to contribute a few hundred cores.

In this paper, we present a different approach taken to process the data from LIGO's first observing run in 2015 and 2016.  We built an infrastructure which utilized remote streaming from a centrally located data store to individual PyCBC jobs.  This allowed us to remove the requirement for a local copy of the data, significantly increasing the number of sites LIGO could utilize.  In 2015, this infrastructure was accessed using standalone transfer clients; in 2016, the approach was improved to include CernVM-Filesystem (CVMFS) to provide POSIX-based access.

This infrastructure has evolved from including a basic data access service at Nebraska for LIGO to a data management ecosystem, involving a series of caches, multiple data sources, and POSIX-based access with a broad range of potential applications.  Section \ref{background} outlines the state-of-the-art in the OSG prior to this work and the constituent pieces of our solution.  Sections \ref{implementation} and \ref{results} explains the two phases of the implementation and the resulting performance, respectively.  Finally, in Section \ref{conclusion} we discuss how we plan to evolve this system into a flexible, multi-tenant, global data management system.

\section{Background} \label{background}


Historically, the OSG has focused on providing a high-level computational services such as the dynamic creation of batch systems across multiple sites utilizing the pilot model \cite{sfiligoi2009pilot}.  In the pilot model, a central factory performs remote submission of batch jobs (\textit{pilots}) to a known set of resources.  These pilots are launched by the site batch system and actually consist of another layer of worker node software; this worker node software connects to a larger multi-site resource pool.  OSG utilizes GlideinWMS \cite{sfiligoi2009pilot} to perform this function.

Compared to the GlideinWMS service, OSG's offerings in data access are relatively low-level: they have focused on allowing remote access to a filesystem \cite{allcock2005globus}; organizations have to subsequently build data management systems \cite{rehn2006phedex, garonne2014rucio, chervenak2008wide} on top of the remote access.  Attempts to recreate these large-scale experimental data management solutions for general use have been unsuccessful \cite{levshina2014irods}.  These are necessarily complex - they must track file location, manage space usage, implement robust site-to-site transfer agents, and have operational procedures in place to handle data loss or inconsistencies at sites.  This last item is the most difficult: data loss is treated as an unexpected event yet, over dozens of filesystems (some large and well-run; some small and run by grad students), becomes relatively regular.

Hence, site admins must be responsive and take part of operations for these ``traditional" data management systems to function.  This drives the cost to the organizations that own the distributed architectures -- and excludes opportunistic users such as LIGO from effectively using large-scale data management on the OSG.

On the other hand, one of the key tools utilized on the OSG is CernVM-Filesystem (CVMFS) \cite{buncic2010cernvm}, a global filesystem used to distribute publicly-available software across the planet.  Groups write their software into a CVMFS \textit{repository server} and the software is distributed via HTTP over a large-scale content distribute network (CDN) based on a multilevel series of hierarchical caches.  Unlike the data management systems, data loss - cache eviction - is expected, cache failure is recoverable via failover, and sites do not even need local infrastructure to utilize the system.  However, the existing CDN was oriented toward working set sizes of typical of large software stacks (around 10GB) - insufficient to meet LIGO's data needs.  Further, to achieve scalability, CVMFS defaults to having all data public.

As part of previous work, we introduced extensions to the CVMFS software to efficiently publish much larger datasets kept on a single storage element instead of the single repository server.  Further, the OSG deployed a new series of caches with the \textit{StashCache} infrastructure \cite{weitzeldatafederations, derek_weitzel_2017_377033}; this hardware allows efficient distribution of data when the working set size is up to 10TB.

Underlying the StashCache software, XRootD \cite{dorigo2005xrootd} is software suite to provide efficient data access to an underlying filesystem.  Multiple \texttt{xrootd} daemons can be clustered together into a a distributed data access infrastructure.  When multiple sites cluster to provide a single namespace, we refer to this as a \textit{data federation}.  The StashCache caching infrastructure fronts a data federation for OSG users; StashCache provides necessary scalability so OSG users do not need to provide scalable storage to be in the data federation.

\begin{figure}[ht]
\centering
\includegraphics[width=0.45\textwidth]{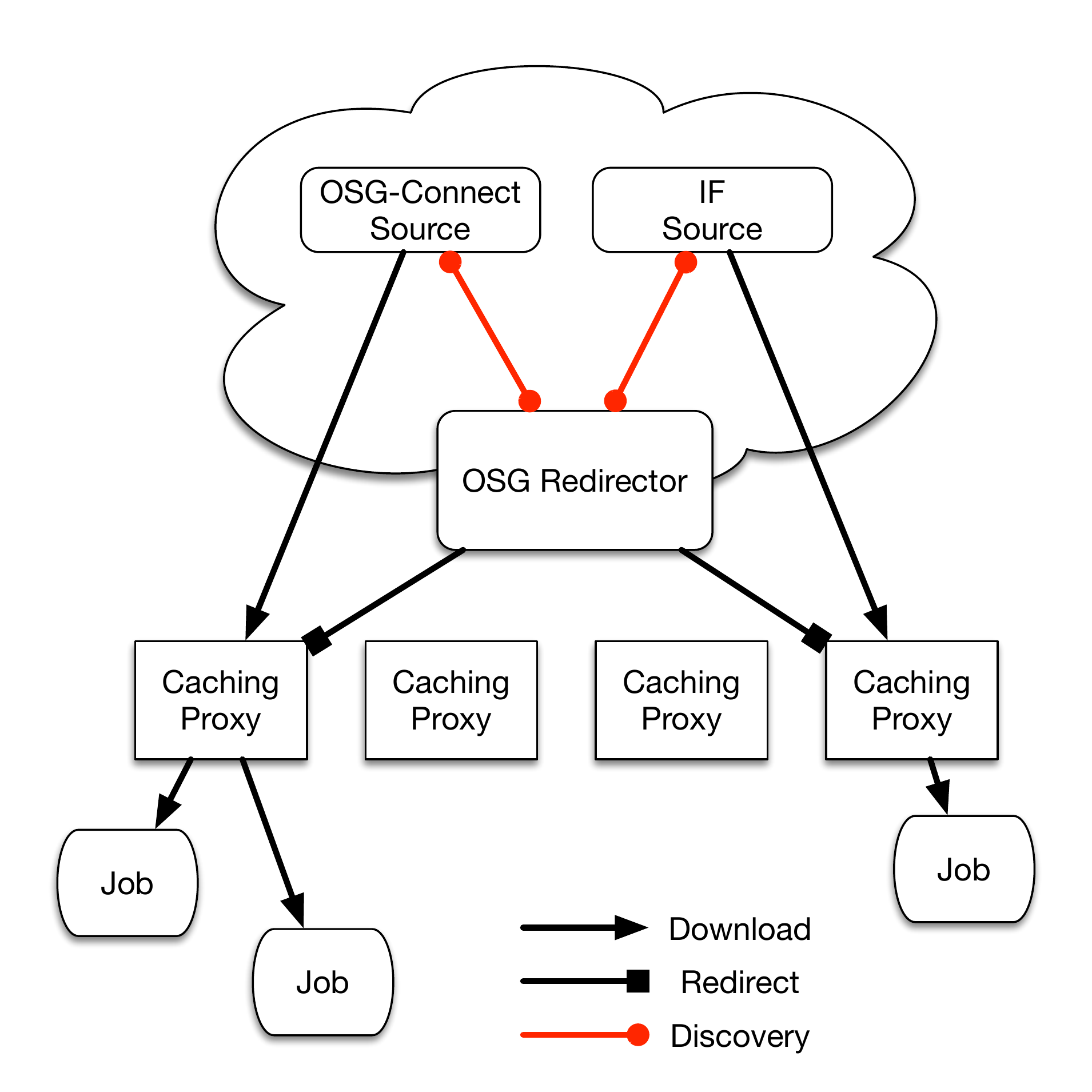}
\caption{Architecture of StashCache}
\label{fig:stashcachediagram}
\end{figure}

Figure \ref{fig:stashcachediagram} shows the architecture of the StashCache, including both the data federation and OSG-run caches.  Jobs, at the bottom, request data from the caching proxies, which are running XRootD.  If the caching nodes do not have the data, the caching nodes query the OSG's XRootD redirector to determine the data source.  At the top, multiple data sources provide the authoritative source for data to the caching nodes.  The data is streamed to the job as it arrives at the caching nodes.

The caching proxies are geographically distributed across the U.S.  The jobs use software such as CVMFS or \texttt{stashcp} that determine the nearest cache.

\section{Implementation} \label{implementation}

\subsection{PyCBC Data Needs}

The PyCBC workflow consists of approximately a hundred thousand jobs for 
each day's worth of recorded LIGO data; the total need is driven by
various aspects of the science (for example, enough data must be analyzed to measure the statistical 
significance of detection candidates) and the computational aspects (re-running failed jobs,
re-submitting incorrectly constructed workflows, or running a workflow
on a new version of the scientific application).  The workflows themselves are managed using the Pegasus Workflow Management System \cite{deelman2005pegasus}.  Each job in the workflow reads one
or two \textit{frame files}, containing the calibrated output of the LIGO 
detectors.  Each file is approximately 400MB and represents 4,096 
seconds of detector data.  The average job in a PyCBC workflow has an execution time between
one hour and tens of hours. Jobs typically run on single cores. Hence, the average sustained
input data needs is 1Mbps per job.  The ``Observation Run 1" (O1) data from 2015-2016 is approximately 7TB.  The ``Observation Run 2'' (O2) data, starting in late 2016 and ongoing at time of writing, is 3TB.  Hence, the total working set size of any individual PyCBC workflow will be less than 10TB.

For science-quality results covering a science run, the data will be re-read approximately 200 times and set of workflows needed will consume several million CPU hours.

\subsection*{Centralized OSG Implementation}

PyCBC workflows are submitted to a single HTCondor scheduler which is hosted at Syracuse University; this scheduler is part of a local HTCondor-based pool of computational resources.  To integrate with the
OSG, we reconfigured this scheduler to ``flock'' \cite{condor-flock} to an OSG-run HTCondor pool.  This causes the scheduler to act as if it were a part of both
resource pools, allowing it to simultaneously and transparently run jobs in both with no changes necessary for the LIGO user.  The OSG pool consists of HTCondor worker nodes launched inside independent batch system across the OSG.  Hence, the LIGO user sees the aggregated resources as a single coherent ``site,'' despite it being virtualized across the grid.  We utilize a central CVMFS repository \cite{bockelman2014oasis} to distribute the LIGO software, providing users with uniform access to necessary software and small amounts of auxiliary data.

The original setup of the data distribution for LIGO was purposely simple: just as the OSG pool appears to the user as a single large computational resource, we would run a central storage service and use it as if it were a large shared filesystem connected to a large cluster. 

We utilized an instance of the Hadoop Distributed Filesystem (HDFS) \cite{bockelman2009using} at Nebraska that was deployed for the CMS experiment; while LIGO needed only a small percentage of the filesystem's total volume, we would take advantage of existing infrastructure to scale transfers.  Nebraska has deployed 12 data transfer nodes (DTNs) on top of HDFS and has 100Gbps connectivity to the R\&D networks in the US.  Each DTN has a single 10Gbps interface and the LVS \cite{zhang2000linux} software is used to provide a single, load-balanced IP address.  This means a remote user could theoretically saturate the entire outbound 100Gbps network connection using the single endpoint.

LIGO's LDR application would transfer data into a storage element at each ``LIGO site'' (here, Nebraska and Syracuse) using GridFTP.  At workflow creation time, Pegasus was instructed to submit to either the OSG HTCondor pool or the Syracuse HTCondor pool, and to treat the HTCondor cluster as able to access the respective GridFTP server.  At job startup, the input data would be downloaded from the configured endpoint via GridFTP directly to the worker node - regardless of the physical location of the worker node.  Jobs would read from Nebraska in the same manner regardless of whether they were physically running in Nebraska, Caltech, or at any other OSG site.  Given that the \textit{average} transfer speed needed per core was 1Mbps out of a total 100Gbps available, this setup was expected to scale across the 10,000 cores we estimated could be available to LIGO.

\begin{figure}[ht]
\centering
\includegraphics[width=0.45\textwidth]{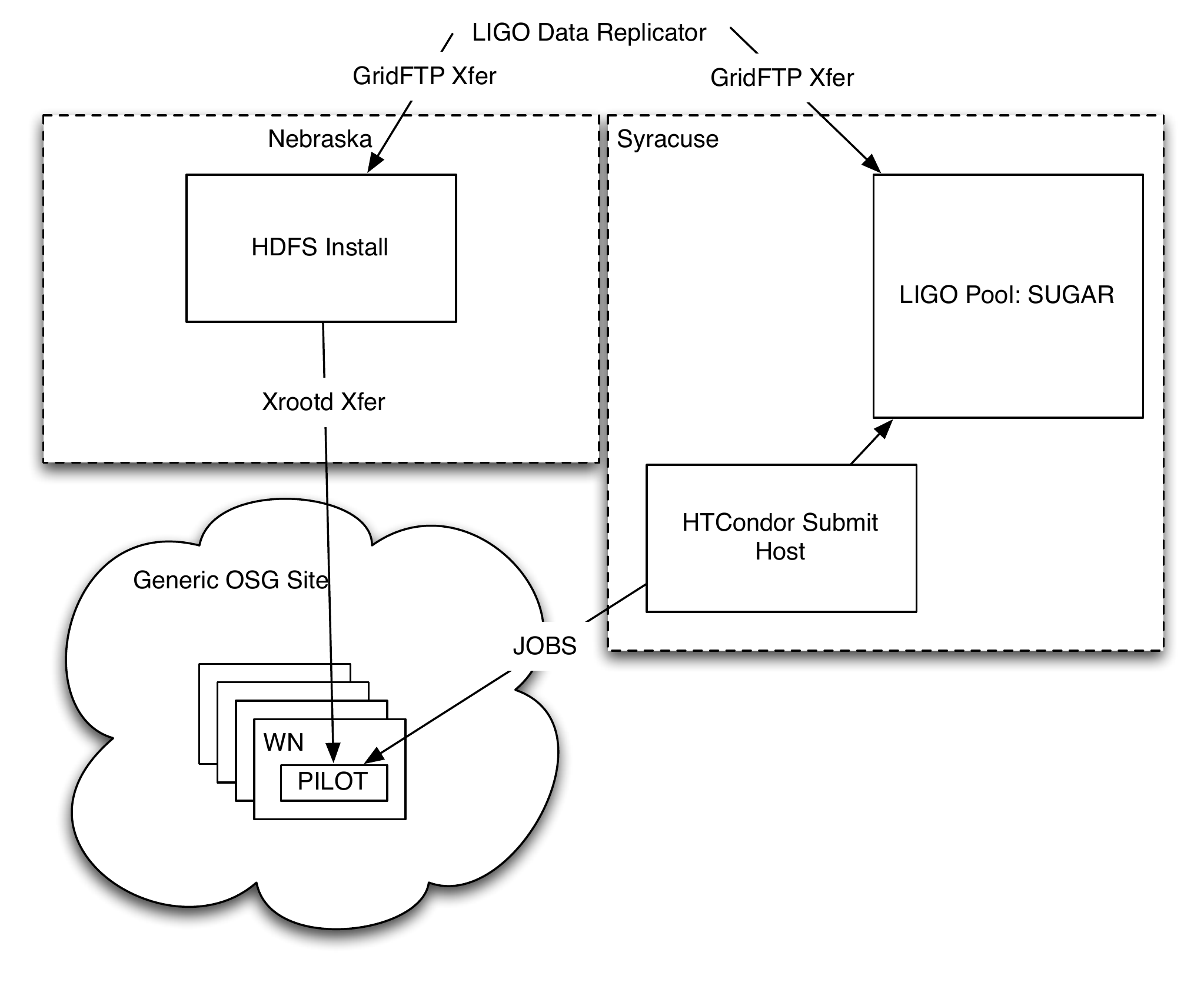}
\caption{Centralized storage deployment of O1 data for processing on the OSG.}
\label{fig:gridftp_architecture}
\end{figure}

As soon as LIGO was able to ramp up to more than 5,000 simultaneous jobs, a few data access issues were evident.

\textbf{Workflow partitioning}: Note, in Figure \ref{fig:gridftp_architecture}, the LIGO user submitting from Syracuse could submit to on-site resource or OSG.  In the initial implementation, they could not use both: the input URLs generated by Pegasus could either refer to the Syracuse storage (NFS-mounted) or Nebraska's (GridFTP).  In response, the Pegasus added the capability to have a prioritized list of input URLs; instead of reading from one source (failing if the file is missing), the jobs would read sequentially through the list of potential sources.  The same job could run on either resource - no longer partitioning the pools and potentially increasing reliability, as jobs could fall back to an alternate source in case of transient failure.

\textbf{Workflow ramp-up}: GridFTP (with the HDFS backend used at Nebraska) consumes approximately 128MB RAM per connection due to the embedded Java virtual machine in the HDFS client and GridFTP starting a separate process per connection.  When workflows were submitted to HTCondor, GlideinWMS began to fill the pool with empty slots and HTCondor would rapidly launch new jobs.  In turn, many new jobs would connect to the GridFTP servers simultaneously, exhausting available memory on the server.  Although the endpoint could manage the steady-state transfer load, it was woefully underpowered for ramp-up.   To avoid this, we utilized the HTCondor scheduler's job startup rate-limiting feature.  Through trial-and-error, we found a job startup rate of 1.5Hz was manageable \textit{if} the individual 400MB file transfers completed under 5 minutes.  Unfortunately, large computational sites with poor TCP throughput to Nebraska (potentially caused when all the site's worker nodes shared an underperforming NAT) would still cause memory exhaustion.  To better protect the service, we developed and deployed a Globus GridFTP extension to limit the number of concurrent transfers per user on each server: when LIGO had too many ongoing connections, the server would simply refuse to start additional transfers.  This would causing queuing or, eventually, a transfer failure (instead of service failure).

\textbf{Scalability}: As the number of LIGO users and available cores increased, balancing job startup rates became an impossible task.  We have little control over how quickly pilots would launch on remote sites' batch queues; if the rate-limits were set too low, the pilots would idle and waste available CPU resources.  Set too high, the concurrency limits would cause payload jobs to fail.  Since the ``correct" value varied by site and the mixture of available sites differed hourly, the data access system led to significant waste.

This caused us to switch from using the Globus GridFTP implementation to access central Nebraska storage to using the XRootD daemons. The XRootD server has a lighter resource footprint and is designed to handle thousands of low-throughput connections; GridFTP is better suited for dozens of high-throughput connections.  XRootD uses a multi-threaded model, meaning each client connection did not require a full JVM but only a few kilobytes of memory for state.  Hence, switching to XRootD moved the bottleneck from DTN memory limitations back to underlying storage performance.

\subsection{Non-OSG Resources}

Despite LIGO gaining experience, efficiency, and additional opportunistic sites on the OSG, more core hours were needed to complete the PyCBC analysis.  We added TACC's Stampede resource to the available resource pool, utilizing an XD allocation won by the Syracuse PI.    Stampede required us to solve three unique problems:

\textbf{Lack of CVMFS}: CVMFS has been a powerful tool for distributing software globally; unfortunately, it is admittedly specific to the grid community.  It is not available on Stampede: to work around this, we utilize \texttt{rsync} to synchronize select CVMFS repositories onto a scratch filesystem on TACC.  While, in general, the Lustre filesystem at TACC provides less performance for software distribution than CVMFS (and updates occur less frequently), the LIGO software stack is sufficiently lightweight to avoid performance bottlenecks.

\textbf{Input data access}: While Stampede worker nodes provide access to the external network, we had concerns about scalability and performance of serving frame files from offsite.  We utilized the Globus file transfer service \cite{foster2011globus} to make a copy of the O1 data after data-taking was complete.  This sufficed as a ``one-off" and avoided the data management difficulties noted in Section \ref{background}.  The input filenames for the Stampede frame files were added to the list of potential input sources in Pegasus.

\textbf{Lack of scalable grid interfaces}: Stampede provided an API for remote submission of jobs based on Globus GRAM \cite{foster1999globus}.  We found this to initially be a poor match for our pilot infrastructure: the pilots could utilize only a single node at a time and our allocation was large enough to run on a few hundred nodes.  We would hit limits on the number of allowed batch jobs at Stampede before we could fully utilize our allocation.  Hence, we developed a new startup script that utilized the \texttt{srun} tool to launch an independent pilot on each node of a multi-node job.  That allowed a single batch job to run on an arbitrary number of Stampede nodes; after trail-and-error, we settled on 1024 cores per batch job.

\*

Put together, toward the end of the O1 analysis, we were able to add up to 10,000 Stampede cores into our resource pool and saw peaks of 25,000 available cores to a single PyCBC workflow.

After the initial O1 analysis was complete, we started adding non-US sites, particularly those from the VIRGO collaboration, to the resource pool.  VIRGO computational sites utilize the shared EGI infrastructure.  EGI is analogous to an European-based OSG; we previously developed the ability to submit pilots across both OSG and EGI infrastructures for the CMS experiment.  To add VIRGO resources to the pool, we only needed to add the service hostnames of any VIRGO-only sites not already used by CMS.  For data access, we used the same approach as OSG sites: remote streaming of data from Nebraska to worker nodes.  In the case of non-US sites, the difference was the network links involved as transfers were often transatlantic.

\subsection*{Distributed CVMFS-based Implementation}

The O1 experience showed that we could sustain 20,000 running cores of PyCBC using a centralized data access architecture.  We could stream frame files directly from storage to the worker nodes, even when those worker nodes were on a different continent.  However, there were open questions of efficiency - how long did jobs wait for input versus total runtime? - and impact on the shared WAN infrastructure.  Maintaining a list of potential input filenames (one for CVMFS, one for Syracuse, one for Stampede) for each logical file was error-prone and not a scalable operational process.  Finally, the data access setup was largely usable only by LIGO pipelines built with Pegasus; it worked well for the PyCBC team but not for others.

For data access in O2, we had additional goals
\begin{enumerate}
\item \textbf{POSIX-based access}: Instead of requiring users to invoke esoteric file transfer utilities to download input, the data files should be available via the traditional POSIX interface.
\item \textbf{Use of available local storage resources}: If either cache-based or filesystem-based storage resources were available to LIGO at the computation site, these should be used instead of the WAN.
\item \textbf{Uniform namespace}: The LIGO frame files should be accessible via the same filenames at all sites in the OSG resource pool, avoiding the need for site-specific lists of filenames.
\end{enumerate}

All three goals were met by utilizing the StashCache infrastructure and its CVMFS extensions.  We now publish the frame files to a CVMFS repository, \texttt{ligo.osgstorage.org}, after being copied to Nebraska by LDR.  CVMFS - as a FUSE-based Linux filesystem - provides POSIX access, meeting our first goal.

By default, all files in CVMFS are public: this is a necessity, given the use of HTTP caching for the CDN.  LIGO frame files are not public and access should be restricted to collaboration members.  Hence, the LIGO CVMFS repository must enable the ``secure CVMFS'' mode.  Secure CVMFS uses X509 certificates \cite{welch2004x} to authenticate and authorize a user to view the namespace and access data.  The client certificate is copied from the user process's environment by the CVMFS client.  If the access control list, distributed as part of the repository, allows access to the DN, then the requested file is served from the worker node's local cache.  If the data is not in the local cache, the user's certificate and key is used to secure a HTTPS connection to request the data.

The use of HTTPS for data access implies the existing CVMFS CDN (implemented using HTTP caches) is not used for data distribution; however, the CDN is used to distribute the filesystem namespace.  The latter - containing only filenames, directory structure, and file size - is not considered private data.

As with O1, the Nebraska DTN endpoint was used to distribute the data.  For O2, we switched the protocol to HTTPS as \texttt{xrootd} supports both HTTPS and xrootd protocols.  As the HTTPS connection from CVMFS is authenticated with the user's client certificate, we again apply the repository's authorization rules at the Nebraska DTN.  Access control is thus applied twice: once for the access to the local cache and again for access to the remote server.  The latter is necessary to protect against malicious CVMFS clients.

Toward the second goal, we have extended the StashCache infrastructure to provide an authorization layer.  Trusted caching proxy servers, shown as the middle layer in Figure \ref{fig:stashcachediagram}, are configured to only allow authorized users (periodically updating the access control lists from the local CVMFS mount) access to the LIGO frame files in the site-level cache.  On a cache miss, the caching proxy connects to the Nebraska origin, authenticating with a local X509 client certificate, and downloads the missing file.  This last authentication step implies that only proxies trusted by LIGO can participate as caches.  This limitation appears unavoidable, but does allow LIGO sites to ``bring their own'' storage resources.

Finally, CVMFS clients are then configured to point at all known caches in addition to the Nebraska-based source.  When first mounted, the CVMFS client will use a GeoIP API to determine a preferred ordering of the potential sources.

If the LIGO-trusted site does not provide a cache server but has a LDR-managed - or Globus-managed - set of frame files on a shared filesystem, we use a CVMFS \textit{variant symlink}.  The derefenced value of the symlink is managed by the CVMFS client configuration.  The symlink at \texttt{/cvmfs/oasis.opensciencegrid.org/ligo/frames} points at \texttt{/cvmfs/ligo.osgstorage.org/frames} by default. However, the local site admin may change the client configuration so the symlink dereferences to \texttt{/mnt/ligo\_nfs/frames} at their local site.  This variant symlink helps achieve our third goal: frame files appear under the same directory whether the CVMFS-based or shared filesystem-based distribution method is used.

\section{Results and Feedback} \label{results}

\subsection*{O1 Results}

\begin{figure}[ht]
\centering
\includegraphics[width=0.45\textwidth]{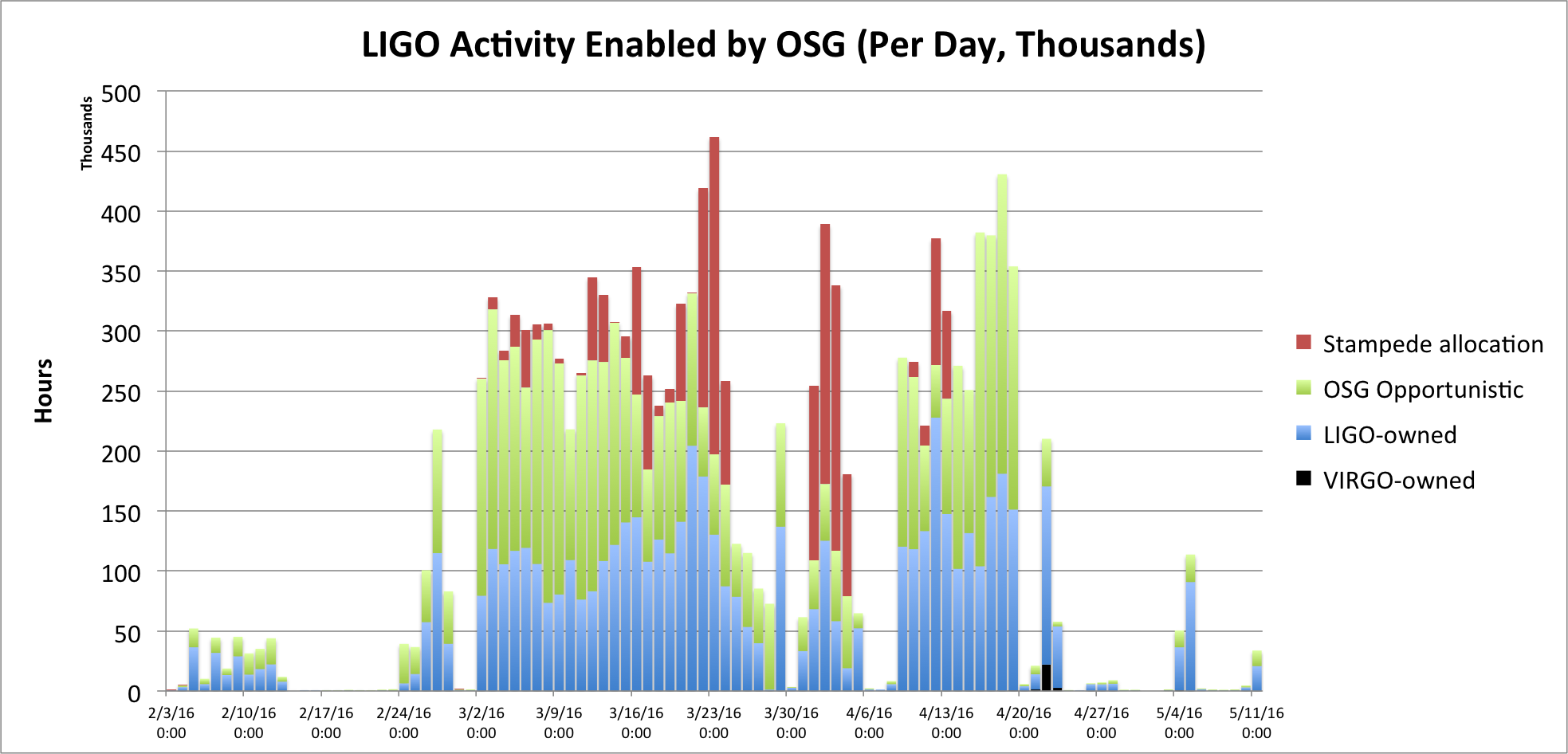}
\caption{LIGO core-hours enabled by LIGO during early 2016.}
\label{fig:ligo_o1_hours}
\end{figure}

The first significant usage of OSG by LIGO for the PyCBC workflow was to process the O1 data.  This occurred in late 2015 and for a second pipeline in early 2016.  Total time used was approximately 4 million hours for 2015 and 13.8 million hours in 2016; about 20 unique computational resources were used for this run.  Figure \ref{fig:ligo_o1_hours} shows the CPU hours utilized in early 2016.

Note that peaks occur around 400,000 hours per day, which is approximately 16,000 jobs continuously occupying CPU cores.  LIGO reports that this project contributed approximately 18\% of their overall O1 computing.  Included in the 13.8M hours in early 2016 is a 2M SU allocation at Stampede; as OSG is composed of many heterogeneous clusters, comparing an ``OSG hour" against a Stampede ``service unit" is difficult.  The OSG maintains a basket of benchmarks  based on the scientific applications of user payloads; using this metric, the ``average'' OSG CPU is 3.147 while Stampede's CPUs rates a 4.599.  Hence, 13.8M total hours is comparable to a 10M SU allocation on Stampede.

\subsection*{O2 File Distribution Methods}


\begin{table}[ht]
\begin{tabular}{|l | c | c | c|}
\hline
\textbf{Site} & \textbf{Secure CVMFS} & \textbf{LDR} & \textbf{Other} \\
\hline
Comet & \checkmark & & \\
Fermilab & \checkmark & & \\
Georgia Tech & \checkmark & \checkmark & \\
Caltech/Louisiana  & \checkmark & \checkmark & \\
Nebraska & \checkmark & \checkmark & \\
NIKHEF & & \checkmark & \\
Omaha & \checkmark & & \\
Michigan & \checkmark & & \\
Polish VIRGO & & \checkmark & \\ 
Syracuse & \checkmark & \checkmark & \\
UCSD & \checkmark & & \\
Wisconsin & & & \\
TACC & & & \checkmark \\
\hline
\end{tabular}

\caption{Availability of data at processing sites}
\label{tbl:dataavailability}
\end{table}

At the time of writing, LIGO's O2 is ongoing, hence we do not know the aggregate PyCBC CPU usage.  Rather, we will focus on how we improved data access for O2.

Table \ref{tbl:dataavailability} shows the data availability for the largest sites running the O2 data processing.  The sites are listed, and checkmarks signify how the data is available to processing at that site.  Approximately half of the sites use CVMFS to provide the LIGO data.  Some sites, such as Georgia Tech, NIKHEF, and the Polish Virgo site are part of the Ligo Data Grid, and therefore do not need CVMFS to access the data.  Sites \textit{not} in this table are smaller and typically use the remote streaming method from O1.

TACC is an example of out-of-band data distribution.  The LIGO collaboration copies the entire LIGO data set to TACC using Globus.  This requires human intervention and planning in order to coordinate the data movement.



\begin{table}[ht]
\begin{tabular}{|l | r| r |}
\hline
\textbf{Access Method} & \textbf{Time} & \textbf{Transfer Speed} \\ \hline
CVMFS & 13.7s & 31 MB/s \\
XRootD Cached & 9.38s & 46 MB/s\\
XRootD Remote & 108s & 4.0 MB/s\\
\hline
\end{tabular}
\caption{Measured transfer speeds for data access}
\label{tbl:transferresults}
\end{table}

Table \ref{tbl:transferresults} shows the transfer times for a 437MB file to a Syracuse worker node.  While CVMFS provides a POSIX interface, this comes at a performance penalty.  We believe this penalty is small tradeoff for the improved interface.  XRootD access is measured using the native XRootD transfer tool, \texttt{xrdcp}.  The ``XRootD cached" column is accessing a local Syracuse cache of the LIGO data set, while the ``XRootD remote" column is accessing the Nebraska DTN over the WAN.

We believe the performance penalty for CVMFS-based access is due to three factors: CVMFS will checksum the file as it is downloaded (XRootD does not); in the current implementation, CVMFS performs a HTTPS download for each 32MB chunk of the input (and starts a new TLS connection); and the context-switching penalty from FUSE.

\section{Conclusion and Future Work} \label{conclusion}

Prior to this work, distributing datasets on the scale of LIGO's frame files was a daunting task.  The approach used by large experiments such as CMS or ATLAS required significant operational effort from both the experiment and the local site: this was a non-starter for opportunistic OSG users like LIGO.

For LIGO's O1 run, we were able to successfully utilize tens of millions of CPU hours for PyCBC by using the central resources at Nebraska and modeling the workflow as if OSG was a large site.  For O2, we utilized a cache-based distribution model to decrease wasted CPU and WAN bandwidth, improve use of local storage, and provide an easier user experience for non-PyCBC workflows.

The CVMFS-based model is \textit{not} LIGO-specific but does require the workflow's usage pattern to be cache-friendly and have a working set size of less than 10TB.  It allows opportunistic usage of storage, but requires the experiment to have at least one origin (such as Nebraska) and to have a minimal level of trust with the cache sites.  Local storage is not required, but the workflows can be WAN-intensive if it is not provided.

We believe this model - even with these restrictions - applies to a wide variety of workflows, including other physics experiments with small datasets and the BLAST databases common in genomics.  We plan to utilize it repeatedly in the future.  As O2 finishes in 2017 and the PyCBC analyses ramp up, we hope to deploy additional caches for more scalability.  Looking further forward, we acknowledge that the X509-based authentication infrastructure is relatively unfriendly to users: we hope to investigate HTTPS-friendly alternates such as those built on OAuth 2.0 \cite{oauth2}.  Altogether, we aim to evolve this into a scalable, easy-to-use infrastructure for all LIGO analysis users.

\section*{Acknowledgements}

This material is based in part upon work supported by the 
National Science Foundation under grant numbers PHY-1148698 and ACI-1443047. Any opinions, findings, and conclusions or recommendations 
expressed in this material are those of the authors and do not necessarily
 reflect the views of the National Science Foundation.
 
\bibliographystyle{ACM-Reference-Format}
\bibliography{main.bib} 

\end{document}